\documentclass[compsoc, conference, a4paper, 10pt, times]{IEEEtran}

\usepackage{cite}
\usepackage{amsmath,amssymb,amsfonts}
\usepackage{algorithmic}
\usepackage{textcomp}
\usepackage{xcolor}
\usepackage{booktabs}

\usepackage{graphicx} 
\usepackage{hyperref}
\usepackage[hyphenbreaks]{breakurl}

\usepackage[disable]{todonotes}

\begin{document}

\title{Security Analysis of Chain-FS service\\
\vspace*{20pt} \normalsize  \today{} }
\author{\IEEEauthorblockN{Vanessa Teague}
\IEEEauthorblockA{\textit{Thinking Cybersecurity Pty.\ Ltd.} \\
vanessa [at] thinkingcybersecurity [dot] com}
\and
\IEEEauthorblockN{Arash Mirzaei}
}

\maketitle
\begin{abstract}
    We examine the security of a cloud storage service that makes very strong claims about the ``trustless'' nature of its security. We find that, although stored files are end-to-end encrypted, the encryption method allows for effective dictionary attacks by a malicious server when passwords only just meet the minimum length required. Furthermore, the file sharing function simply sends the decryption passwords to the server with no protection other than TLS.
\end{abstract}

\section{Introduction}
Chain-FS (\url{https://www.chain-fs.com/}) is a file storage and sharing system that claims very strong privacy properties.
This is the way Chain-FS describes itself:
\begin{quotation}``Chain-FS is a secure storage API that allows for the immutable storage of data, as well as providing the data owner with complete control of information sharing, and unprecedented privacy.

Chain-FS is a trust-less architecture, meaning that unlike traditional cloud stores, Chain-FS delivers a platform that says, there is no need to trust us with your data, because we cannot read it, share, or modify it, and this includes our system
administrators.''\footnote{\url{https://www.chain-fs.com/about} Last accessed Mar 31, 2025.}
\end{quotation}

\begin{figure}
    \includegraphics[width=0.5\textwidth]{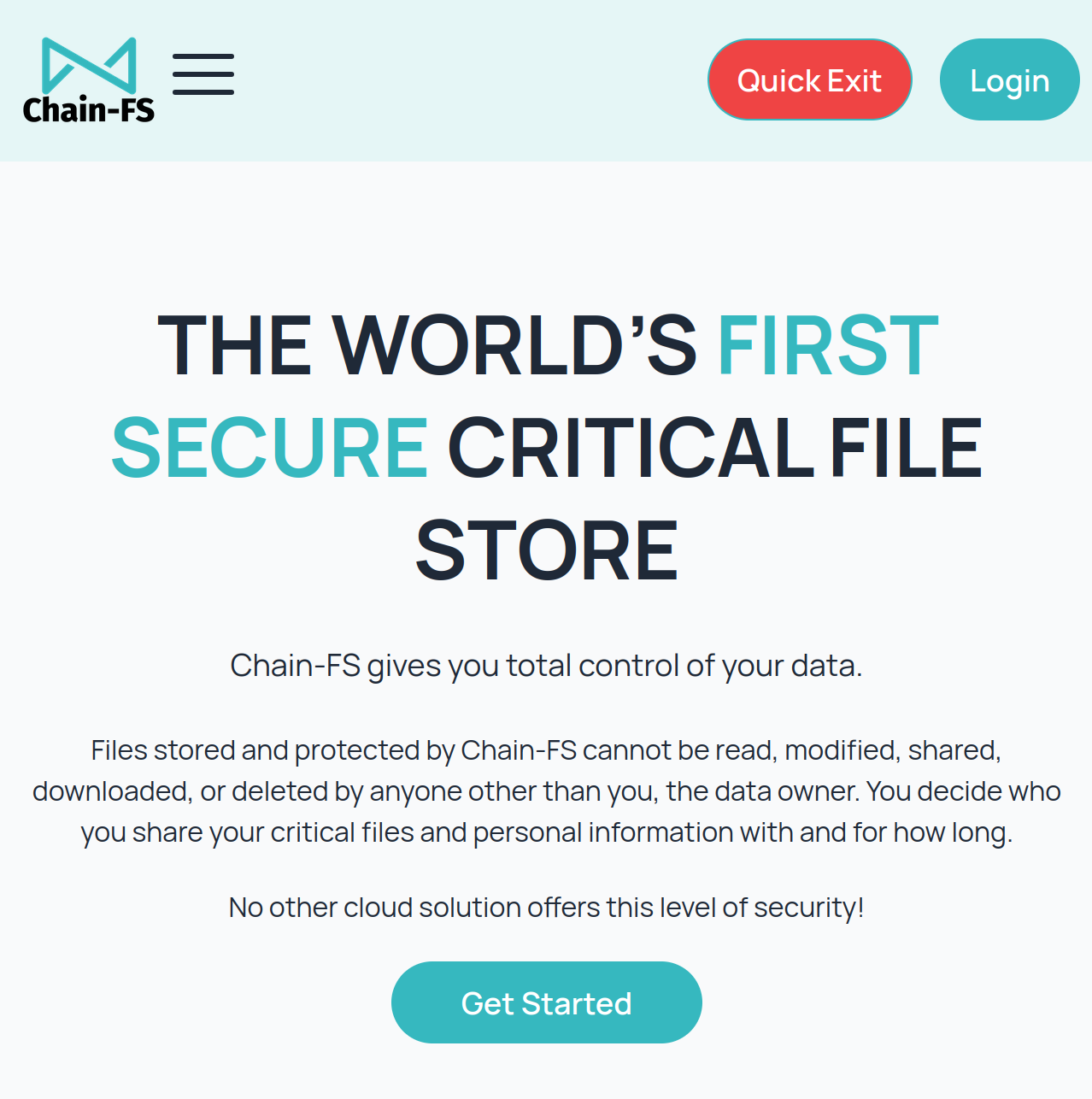}
    \caption{The main landing page of chain-fs.com, as at 31 Mar 2025. Claims that even administrators cannot read the data feature prominently.}
    \label{fig:splash-screen}
\end{figure}

These sorts of claims are made prominently: a screenshot of the main landing page, with similar assurances, is shown in \autoref{fig:splash-screen}.

The purpose of this work is to analyze the security features of Chain-FS and investigate if it is able to provide a trust-less architecture.
We analyzed the JavaScript code executed in the browser, which was accessible through the browser's developer tools.
We identified two main problems.
\begin{enumerate}
\item File encryption can be easily reversed through systematic password-guessing, when passwords meeting the minimal accepted length are used. We were able to iterate through all permitted 6-character passwords on an ordinary laptop in around 30 hours. This means that an attacker with control of the server can read at least some properly-encrypted files. The details and suggested corrections are in \autoref{sec:storeFile}. Since our initial report, the minimum length has increased to 8 characters, which makes no substantial
difference.
\item File sharing simply sends the passwords to the server in the clear (apart from TLS). This undermines the security goals of an end-to-end encrypted service and means that the architecture relies entirely on trusting the server not to decrypt a file that has been shared. It is also vulnerable to any attackers able to subvert the TLS connection, including proxies. The details are in \autoref{sec:fileShare}. 
\end{enumerate}

\autoref{sec:storeFile} describes the implementation of encryption including some technical errors.
\autoref{sec:otherIssues} gives an overview of other issues and questions. \autoref{sec:notification} describes the
disclosure history of this report.

These shortcomings are particularly important given the way this product is marketed. Chain-FS is actively marketed in Australia to survivors of domestic violence. Since Australia has laws for compelling data access from corporations, the promise of a
trustless architecture, with the technical guarantees of end-to-end encryption, may be particularly critical for this cohort.

\subsection{Related Work}
Many other works~\cite{bock2016pwncloud ,niehage2020cryptographic, heninger2023hidden , backendal2023mega, albrecht2024share,    hofmann2024end} have found that services advertising end-to-end encrypted file storage did not achieve their advertised security properties. These related to a variety of issues including metadata leakage and the opportunity for the server to alter files undetectably, as well as breaches of confidentiality and integrity of stored files. Most of them required more subtle attacks than we describe in this paper.

\section{Store File} \label{sec:storeFile}
\begin{figure}[!htp]
\centering
\includegraphics[width=0.5\textwidth]{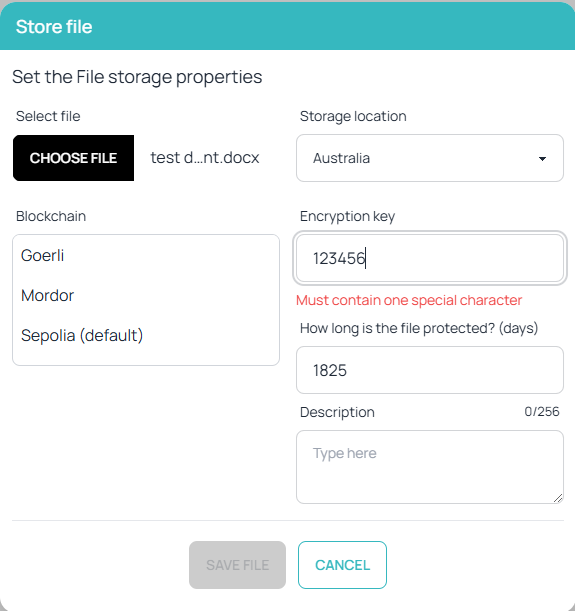}
\caption{Store File Window, before our notification to Chain-FS. Minimum password length is 6.}
\label{fig:storefile}
\end{figure}

\begin{figure}
\centering
\includegraphics[width=0.5 \textwidth]{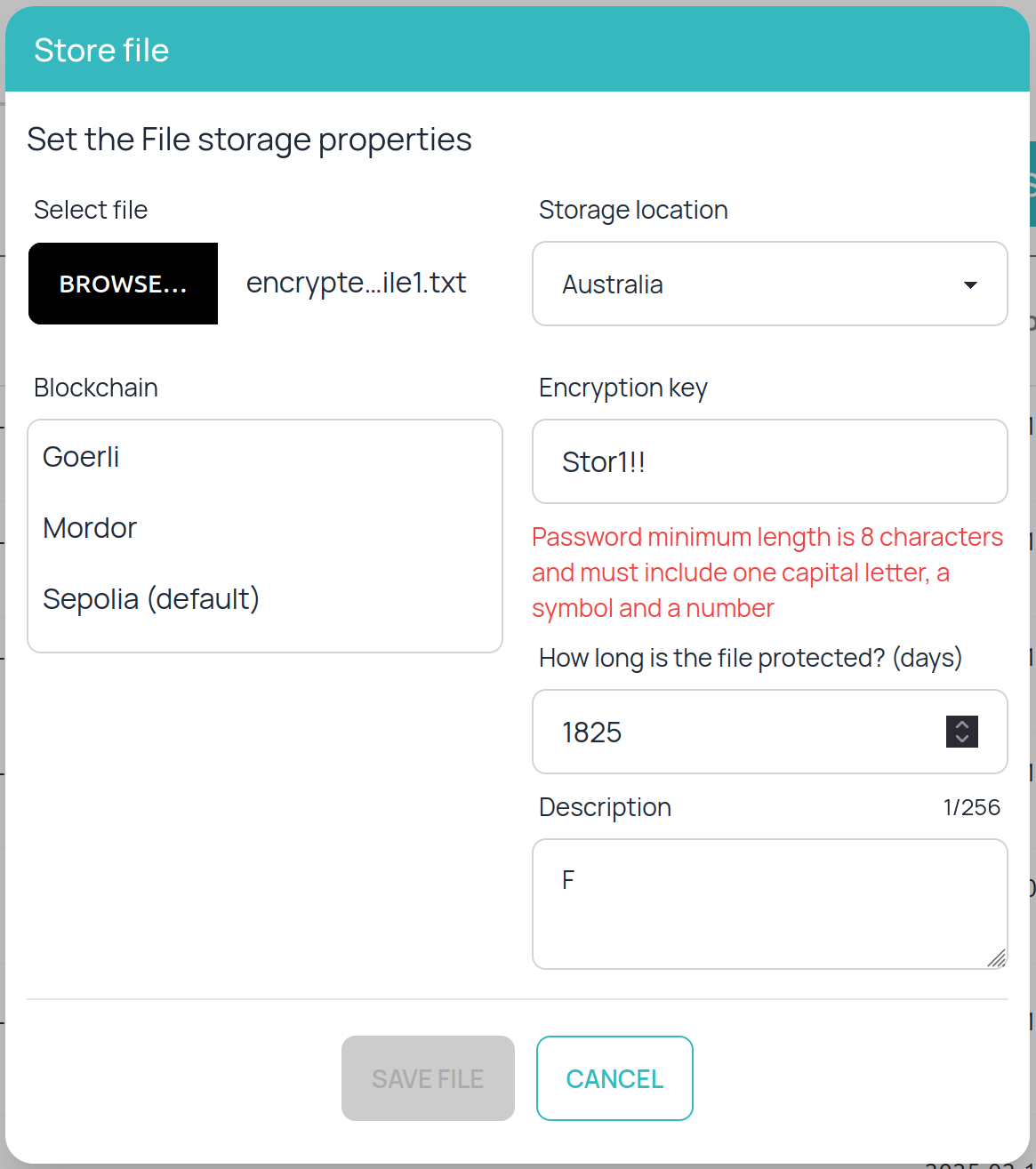}
\caption{Store File Process, more than 90 days after our notification to Chain-FS. The minimum password length has been increased to 8.}
\label{fig:storefile-enc-8-char-pwd}
\end{figure}

\autoref{fig:storefile} illustrates the options that the user must complete when storing a file. \autoref{fig:storefile-enc-8-char-pwd}
shows that the password length has now increased from 6 to 8 characters.

\subsection{How file storage works}
To store a file, the user must provide a password in the `Encryption key' field, as shown in \autoref{sec:storeFile}.
The user-provided key must be between 6 and 32 characters long and must include at least one special character (i.e., a character from the set \{!@\#\$\%\^\&*\}). (Since our
first report to Chain-FS, the minimum password length has increased to 8 characters, but this makes no substantial difference to the analysis presented here. See \autoref{fig:storefile-enc-8-char-pwd}). As a result, users are permitted to use very short keys such as $!!!!!!$. The user-provided key is directly used as the AES key without utilizing standard key derivation methods such as PBKDF2. If the user-provided key is shorter than 32 characters, it is padded with zeros to reach the required length. If the key exceeds 32 characters, an error is generated.

The selected file is encrypted using AES-256 in CBC mode with
PKCS7 padding and an all-zero Initialization Vector (IV), then hashed using SHA-256. Both the encrypted file and its hash value are uploaded to the server. Upon successful storage, the server responds with a unique file GUID. To retrieve the file, the user provides the same password, which is used locally to decrypt the encrypted file downloaded from the server.

\begin{figure*}
\centering
\includegraphics[width=1  \textwidth]{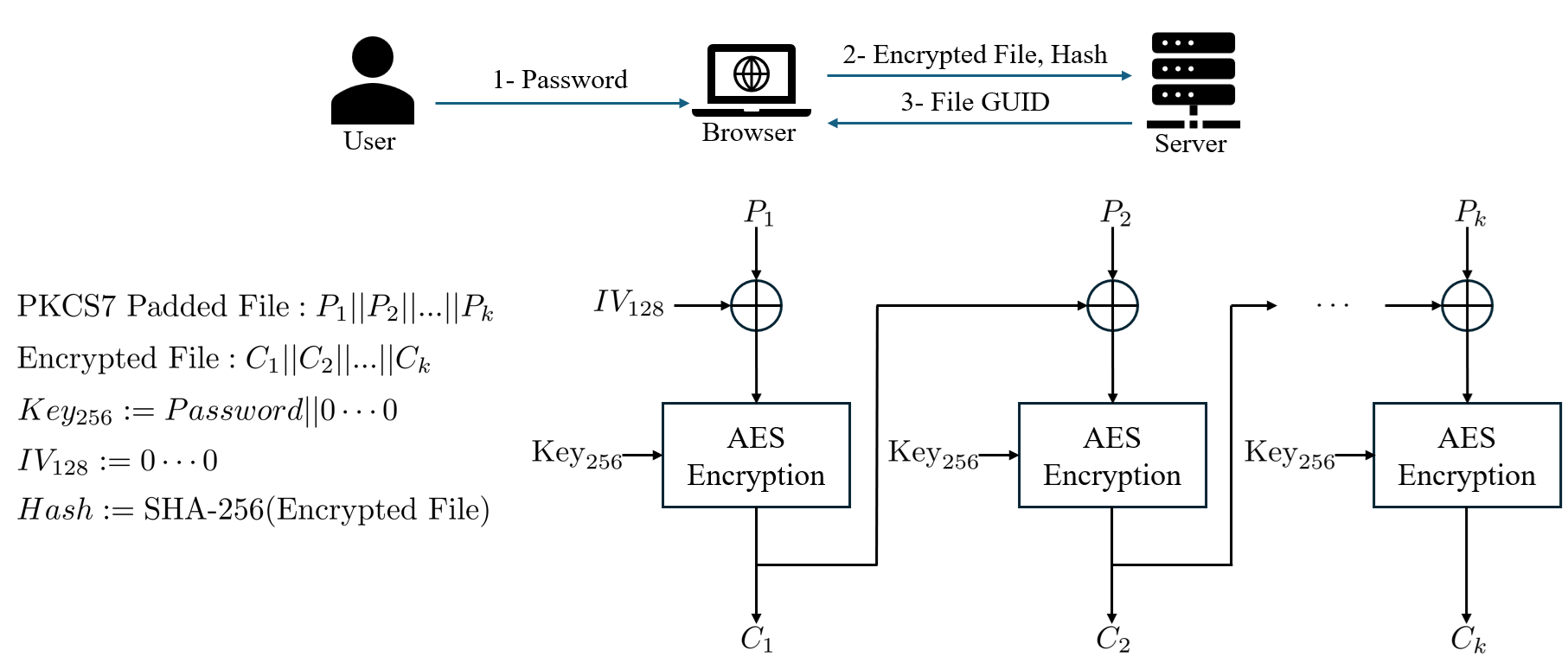}
\caption{Store File Process.}
\label{fig:storefile-enc}
\end{figure*}

\begin{figure*}
\centering
\includegraphics[width=1 \textwidth]{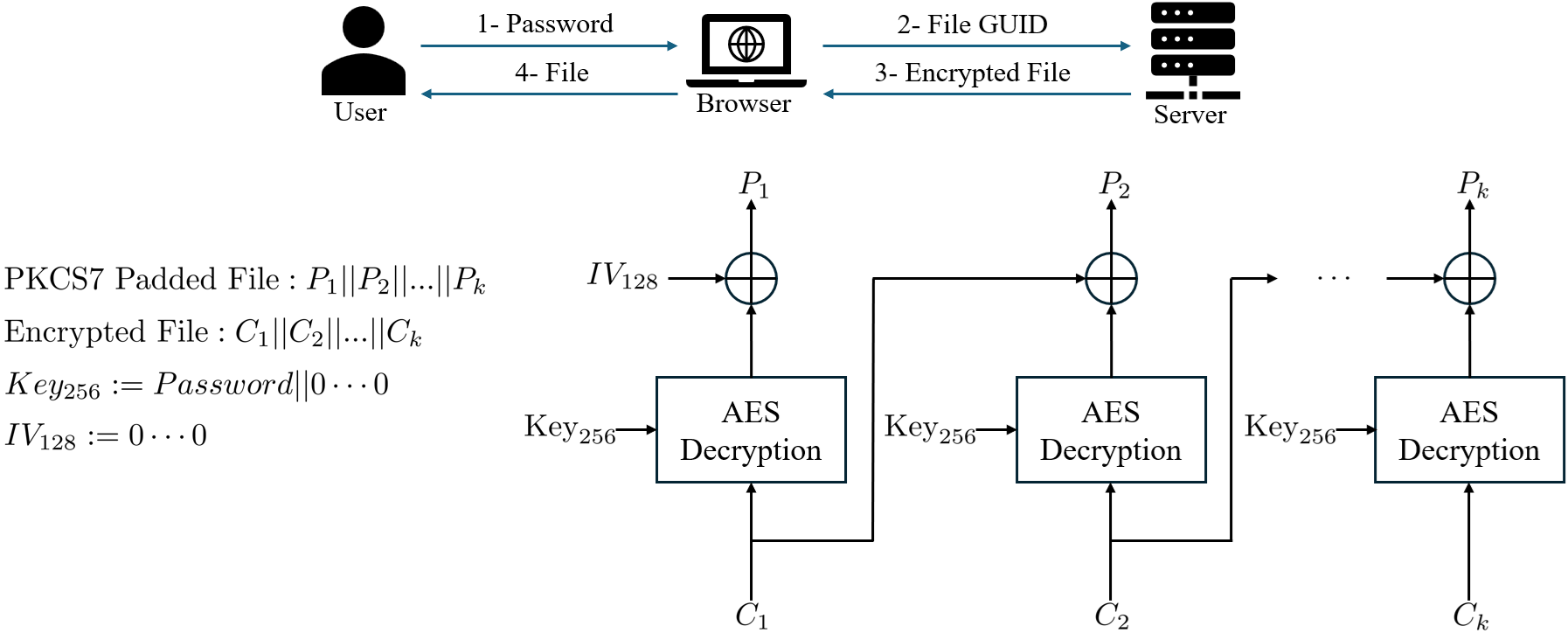}
\caption{Download File Process.}
\label{fig:storefile-dec}
\end{figure*}

\autoref{fig:storefile-enc} and \autoref{fig:storefile-dec} summarize the file store and file download processes.

\subsection{Required cryptographic corrections} \label{subsec:cryptoCorrections}
It is true that file storage and retrieval is an end-to-end encrypted process, with the file encrypted and decrypted on the client side and only the encrypted version sent to the server. However, there are several vulnerabilities in this design that allow a malicious server to extract various levels of information about encrypted files, including complete decryption in some cases. This section
lists the main problems that need to be corrected in the file storage protocol.

    \textbf{Unpredictable Initialization Vector (IV).} Since the IV is an all-zeros vector, encrypting two identical files with the same password will produce identical ciphertexts. Moreover, if two files share the same first $n$ blocks (that is, $16n$ bytes), the ciphertexts for the corresponding $n$ blocks will also be identical, if the same password is used. This compromises the security of the encryption and reveals patterns in the data. To resolve this issue, an unpredictable IV must be used.

    \textbf{Key length and complexity.} The current minimum key length of 6 characters makes brute-force attacks feasible
(and the update to 8 does not substantial impact that). The extra restrictions (one special character, etc) actually reduce the search space. A \emph{randomly generated} password of about 12--14 characters is the minimum required to make exhaustive search attacks impractical.

    However, it is important to note that brute-force attacks are the least efficient way of guessing passwords. Even with a 14-character minimum, users often create predictable passwords (e.g., their pet's name followed by a few special characters).
    Strong emphasis must be placed on the use of properly randomized passwords. Users should be educated about the risks of using simple or reused passwords, as such practices can still allow attackers to decrypt files despite longer key lengths. To address this challenge, integrating password-related fields with trusted password managers that generate and store strong, random passwords could be highly beneficial. Such integration reduces the burden on users to create and remember secure passwords, while ensuring that passwords meet the necessary complexity and randomness requirements. Of course, this introduces a point of trust (i.e. the password manager), but should not necessarily mean that a third party learns the passwords (assuming the password manager runs locally and does not share them).
Alternatively, the user should be asked to generate and remember a high-entropy password. For example, Nextcloud asks for 12 \emph{words}.\footnote{
    \url{https://nextcloud.com/features/\#end-to-end}}

    \textbf{Key Generation from Passwords.} Directly using the user-provided key as the encryption key increases vulnerability to offline attacks. To mitigate this risk, the key should be combined with a unique salt and processed using a standard password-based key derivation function, such as PBKDF2~\cite{kaliski2017rfc} or Argon2~\cite{biryukov2021rfc}. This approach increases resistance to brute-force and dictionary attacks.

\section{Share File} \label{sec:fileShare}
As shown in \autoref{fig:sharefile}, the file sharing process requires the file owner to enter both the original file password and a new sharing password. These are labeled in \autoref{fig:sharefile} as `File decryption key' and `New share password,' respectively.

\subsection{How file share works}
The new sharing password, provided by the file owner, is concatenated with a fixed string (`American Psycho') and hashed twice using Keccak-256 to generate a 32-byte value. This value serves as the private key for the secp256k1 elliptic curve. The corresponding public key is hashed with Keccak-256, and the last 20 bytes of the output are used as the address.\footnote{This is called `PublicKey' in the json, but we use the term `Address' to distinguish it from the full public key.} The original password, the sharing password, and the address are sent to the server, which responds with a new GUID. A link is then created using the GUID and is provided to the file owner.

To share the file, the owner of the file provides the recipient with the link and the new sharing password. When the recipient accesses the link, their browser uses the new sharing password to regenerate the secp256k1 public-private key pair. The private key is used to sign a timestamp, which, along with the signature, is sent to the server. Upon successful verification of the timestamp and signature, the server retrieves the file, encrypted with the new sharing key. The recipient's browser then decrypts the file.

\begin{figure*}
\centering
\includegraphics[width=1  \textwidth]{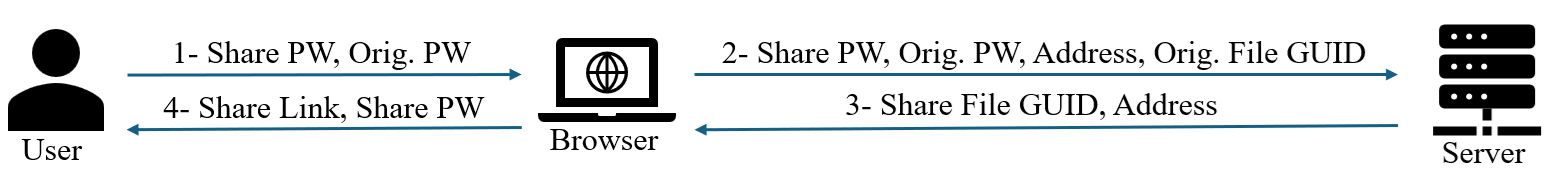}
\caption{Share File Process.}
\label{fig:shareFile-diag}
\end{figure*}

\begin{figure*}
\centering
\includegraphics[width=1\textwidth]{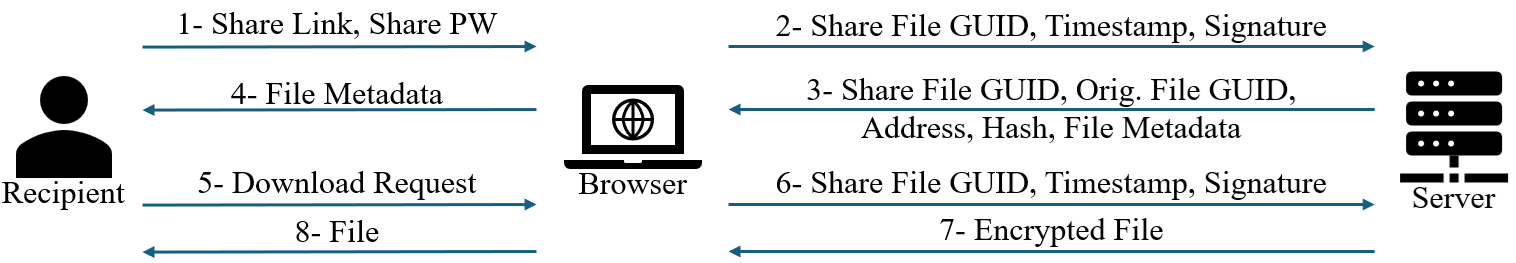}
\caption{Shared File Access Process.}
\label{fig:accessFile-diag}
\end{figure*}

\autoref{fig:shareFile-diag} and \autoref{fig:accessFile-diag} summarize file share and file access processes, respectively.

\subsection{Problems}
There are several issues with this design.
\begin{itemize}
    \item A significant vulnerability in this design is that the browser transmits both the original and the sharing passwords to the server, thus compromising their security. Although TLS protects the passwords from interception by third parties
    (though not TLS proxies or attackers who have compromised the TLS session), there is no protection against the server
    learning the passwords and hence decrypting the files. At this point, it does not seem acceptable to claim a ``trustless'' architecture.
    \item We observed that if the file owner enters an incorrect original password during the file-sharing process, the server can detect the error. However, since the server does not have access to the original password prior to the first sharing attempt, we became curious about how it identifies such errors.

    We guessed that it might be using padding verification, and tested this by generating (by brute-force) a wrong password that nevertheless produced valid padding when decrypted (a block
    whose last byte is 1 or the last 2 bytes are 2). This was accepted, and no other wrong passwords were accepted. This seems like good evidence that valid padding is the acceptance criterion:
    if the padding is invalid, the server returns an error; otherwise, it accepts the password.

    There is a small but non-negligible probability that an incorrect password could result in valid padding after decryption. In such cases, the server incorrectly accepts the password and provides a link to the encrypted file. Although this is clearly not the intended behavior, it also does not seem to lead to plaintext exposure: when the link and the new sharing password are used by the recipient, the decryption is highly likely to produce gibberish.
\end{itemize}

\begin{figure}[!htp]
    \centering
    \includegraphics[width=0.5\textwidth]{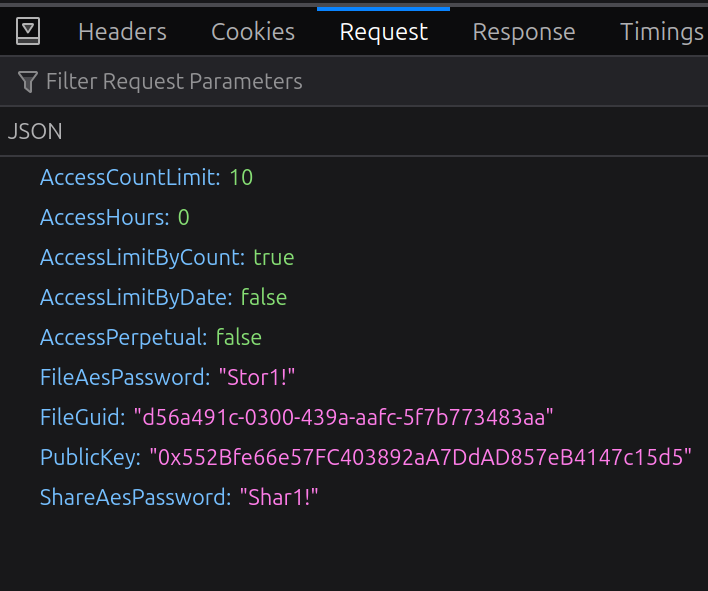}
    \caption{Network traffic for file sharing, showing storage and sharing passwords. Last verified on 30 Mar, 2025.}
    \label{fig:sendPasswords}
\end{figure}

\begin{figure}[!htp]
\centering
\includegraphics[width=0.5\textwidth]{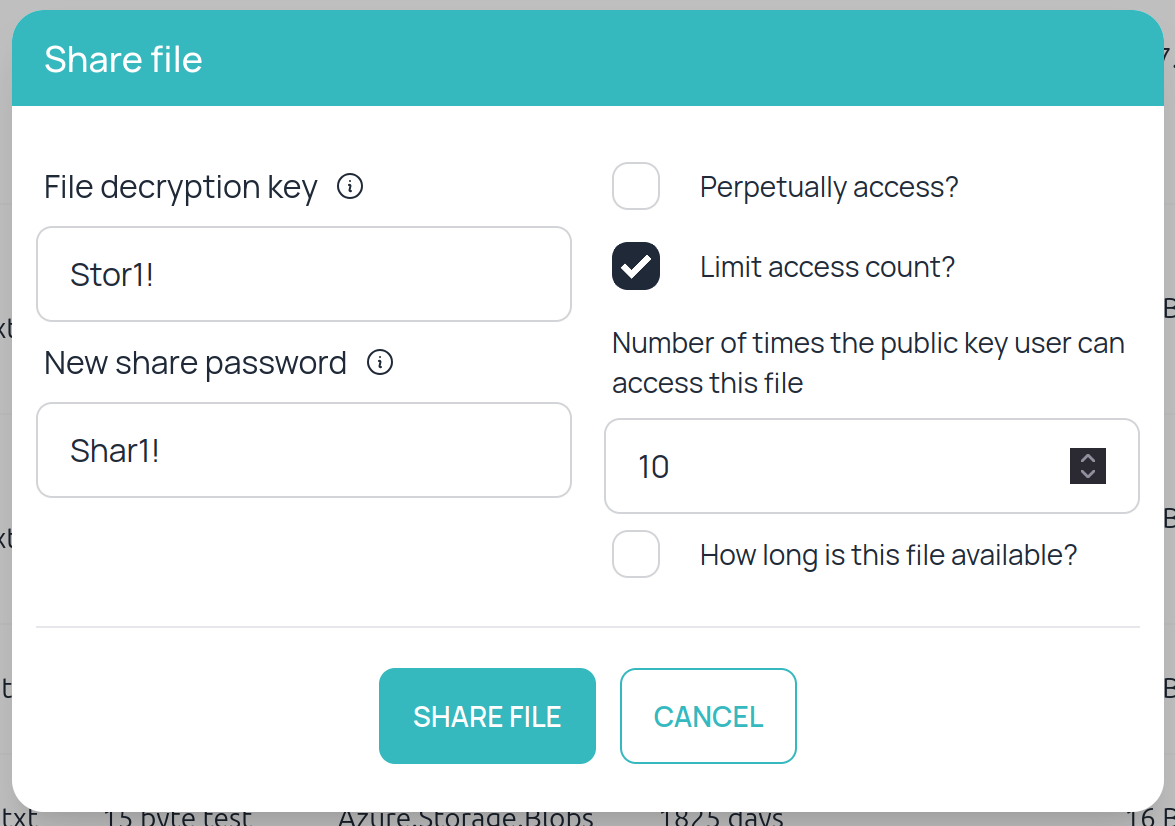}
\caption{Share File Window. Note that the passwords for file storage and sharing match those in \autoref{fig:sendPasswords}.}
\label{fig:sharefile}
\end{figure}

\autoref{fig:sendPasswords} shows the network trace in the browser. Note that both the storage password and sharing password are sent to the server.

\section{Other issues} \label{sec:otherIssues}
\begin{itemize}
\item The Chain-FS FAQ states, ``All files entering Chain-FS are hashed for integrity and to compare to known databases of child abuse material.'' This hash is computed over encrypted data, and consequently will not match any database of illegal images.

Note that it would be a serious mistake to take the hash of plaintext data instead. We are not advocating that alternative, which would expose information about the plaintext. It is probably not possible to combine the guarantees of an end-to-end encrypted service with the promise of scanning for illegal material.

\item The user can choose from the following blockchains: Goerli, Mordor, or Sepolia. However, it is unclear what data is stored on the blockchain or for what purpose. Notably, Goerli, which is an Ethereum testnet, has been deprecated and will no longer be updated. So, it is unclear what will happen when the user selects this blockchain.
\item It is unclear how the data was intended to be placed on the blockchain or why such an approach would be considered necessary.  In any case, it is strongly advised not to store users' data—whether encrypted or not—on the blockchain.  
\end{itemize}

\section{Notification, disclosure, and patches}\label{sec:notification}
The matters described in this report were disclosed to Nansen.io, the owner of Chain-FS, on 13 December 2024. We invited them to work with us to try to fix the issues we had identified, and indicated an expected disclosure time of 90 days. Nansen refused to engage in a discussion about fixing the problems, insisted that we not publish our findings at all, and made an official complaint to the university. We are therefore, unfortunately, unable to express any confidence that the problems described in this report will be corrected. On 30 Mar 2025, we confirmed that the two main problems described in this report are still present, though the minimum password length has been increased to 8 characters.

Nansen states that the version we examined was only a pilot version. However, their website makes no indication of this.

\newpage


\end{document}